\documentclass[prd,aps,twocolumn,showpacs]{revtex4}
\usepackage{epsfig}

\def\bea{\begin{eqnarray}}
\def\eea{\end{eqnarray}}
\def\be{\begin{equation}}
\def\ee{\end{equation}}

\begin{document}
%begin{titlepage}
\voffset 1.5cm 
\preprint{ KAIST-TH 2003/15 }
\title{
Neutrino Mass Matrix in the Minimal Supergravity Model: Bi-large
Mixing with Trilinear R-parity Violation \footnote{ Talk presented by
D.W.Jung  at 2nd International Conference on Flavor Physics (ICFP 2003) Korea Institute for Advanced Study, Seoul, Korea, Oct.6-11, 2003}
%Neutrino Oscillations and Collider Signatures in
%Minimal Supergravity without R-parity: Revisited
}

\author{
Eung Jin Chun$^1$, Dong-Won Jung$^2$ , Sin Kyu Kang$^3$ and Jong Dae Park$^4$ }

\affiliation{
$^1$Korea Institute for Advanced Study, 207-43 Cheongryangri-dong,
Dongdaemun-gu,  Seoul 130-722, Korea\\
$^2$ Department of Physics, Korea Advanced Institute of Science
and Technology,
373-1 Kusong-dong, Yusong-gu, Daejoen 305-701, Korea \\
$^3$School of Physics, Seoul Nat'l University, Seoul 151-747, Korea \\
$^4$Institute of Physics and Applied Physics, Yonsei University,
Seoul 120-749, Korea }

\begin{abstract}
We study the correlation between the neutrino oscillation and
the production/decay of the
lightest supersymmetric particle(LSP) in the context of  the minimal
supergravity model(mSUGRA) without R-parity.
We show how the neutrino masses and mixing which are consistent with
the recent neutrino data can be obtained in this model, and describe
how to probe the model by observing the LSP decay at collider experiments.
It is shown that the generic 1-loop contributions to neutrino masses
 are too small to account for the solar and atmospheric neutrino
oscillations and thus some fine cancellation in tree-level 
contribution is required to severely constrain the viable parameter space.
In most parameter space of mSUGRA, the neutralino or the stau is LSP.
Examining both cases,
 we find that there is a simple correlation between the
neutrino mixing angles and the LSP branching ratios in the small
$\tan\beta$ region so that the model
can be clearly tested in the future colliders.
In the large $\tan\beta$ region, such a correlation is obscured by the large
tau Yukawa contribution which makes it nontrivial to test the model.

%As will be shown, the 1-loop contributions to neutrino masses
% are important to account for
%the solar and atmospheric neutrino oscillations. 
%In mSUGRA, most parameter
%space predicts that the neutralino is LSP, while there is a narrow parameter 
%region which makes the scalar tau LSP. We probe both caeses by calculating 
%the decay rate and branching ratios.
%We find that it is possible to test the model by searching the collider 
%signals for the LSP decay in the small $\tan \beta$ region, while
%it is difficult to probe the model 
%in the large $\tan \beta$ case because  of the large contributions
%from tau Yukawa coupling.
\end{abstract}
%12.60.-i Models beyond the standard model
%13.20.
\maketitle
%\thispagestyle{empty}
%end{titlepage}
%\begin{multicols}{2}
\section{Introduction}
The recent progresses in neutrino experiments has led us to convince
the existence of neutrino masses and flavor mixing.
\cite{Fukuda:1998mi}.
In this regard, of primary interest are to look for
New Physics candidates which are inherited not only with a
natural mechanism to generate the observed neutrino mass matrix and
but also with some other predictions that can be tested in the near
future.
One of the best-motivated models endowing such a property
would be the Minimal Supersymmetric Standard Model (MSSM) with lepton
number
violation via R-parity breaking terms
\cite{Hall:1983id}\cite{hemp}\cite{TRpV}, as
such models may produce lepton flavour violating signals predicted by
the observed neutrino mixing and thus can be tested in future
collider experiments
\cite{Mukhopadhyaya:1998xj}\cite{Diaz:2002ij}.
The particle spectrum in MSSM
depends on the supersymmetry breaking mechanism.
One of the most popular scenarios for supersymmetry
breaking is the minimal supergravity scenario (mSUGRA) \cite{gut}, which
assumes a universal gaugino mass $M_{1/2}$,
a universal scalar mass $m_0$, a universal trilinear coupling $A_0$
and $B_0$ at $M_{GUT}$ scale.
This framework can successfully evade the appearance of dangerous
flavor changing neutral current(FCNC) and is highly predictive because
there exist only 5 independent parameters in the model.
In this work, we do complete phenomenological
analysis on the masses and mixing of neutrinos and lepton number
violating signature of LSP decays in the context of mSUGRA with
trilinear R-parity violation. The neutrino mass matrix is calculated
up to the one-loop order, and it is studied
how the predictions concerned with
neutrino parameters in mSUGRA can be probed from the collider signals.
In almost all parameter space of mSUGRA, the LSP is predicted to be
a neutralino or a stau. In both cases, we calculate
the production cross section, decay rate, and branching ratios of the LSP
to investigate the correlations between neutrino oscillation parameters
and
collider signatures from the various channels of LSP decay.

\section{Neutrino mass matrix and flavor structure of trilinear couplings
in mSUGRA}
Let us begin by writing the superpotential in the basis where the bilinear 
term $L_i H_2$ is rotated away :

\begin{equation}
W_0 = \mu H_1 H_2 + h^e_i L_i H_1 E^c_i + h^d_i Q_i H_1 D_i^c
 + h^u_i Q_i H_1 U^c_i
\end{equation}

\begin{equation}  \label{WRpV}
 W=  \lambda_i L_i L_3 E^c_3 + \lambda'_i L_i Q_3 D^c_3,
\end{equation}
where  $W_0$ is R-parity conserving part, whereas $W$ is R-parity violating part, and we assume the dominance of
$\lambda'_i \equiv \lambda'_{i33}$ and $\lambda_i \equiv
\lambda_{i33}$ ($i=1,2,3$) over the other trilinear couplings of lower
generations and the universality at the GUT scale is imposed. In this framework,
there are 10 free
parameters of the model, i.e. five conventional ones plus five
R-parity violating ones;
$$ m_0,\; A_0,\; M_{1/2},\; t_\beta,\; \mbox{sign}(\mu)\;\; ;\;\;
   \lambda'_{1,2,3},\; \lambda_{1,2}$$
where $t_\beta=\tan\beta$ is the ratio of two Higgs VEVs.
As is well known,  non-universality is developed at the weak scale via
RGE evolution. The RGE in this basis can be found in Ref.~\cite{Chun:1999bq}.
Non-vanishing soft terms are given by
\begin{equation} \label{VRpV}
 V = m^2_{L_i H_1} L_i H_1^\dagger +  B_i L_i H_2 + +B H_1 H_2 + h.c.
\end{equation}
where $B_i$ is the dimension-two soft parameter. 
%The corresponding
%term for the Higgs bilinear is denoted as $B H_1 H_2$. 
We note that non-trivial VEVs for sneutrinos can be generated in this case.
Including 1-loop contributions $V_{loop}$,
we obtain the following
relation which comes from the minimization condition for the scalar potential,
\begin{equation}\label{svev}
 \xi_i = {m^2_{L_i H_1} + B_i t_\beta + \Sigma_{L_i}^{(1)}
          \over m^2_{\tilde{\nu}_i} + \Sigma_{L_i}^{(2)} },
\end{equation}
where $\xi_i \equiv \langle \tilde{\nu}_i \rangle/ \langle H_1^0 \rangle$, and
$\Sigma_{L_i}^{(1,2)}$ are 1-loop contributions generated from $V_{loop}$.
The explicit forms of $\Sigma_{L_i}^{(1,2)}$ are presented in
Ref.~\cite{Chun:1999bq},\cite{phdjung}. Introducing
another variables,
\begin{equation}
 \eta_i \equiv \xi_i -{B_i\over B},
\end{equation}
we can obtain the relation,
\begin{equation}
 {\xi_i -\eta_i \over \xi_i} =
 -{m^2_{\tilde{\nu}_i} \over m_A^2 s_\beta^2}
 {1\over 1+{m^2_{L_i H_1} \over B_i t_\beta}},
\end{equation}
whose non-zero values are due to the neutral scalar loops.
Then, the tree-level mass is presented in terms of $\xi_i$;
\begin{equation} \label{Mtree}
 M^{tree}_{ij} = -{M_Z^2 \over F_N} \xi_i \xi_j c_\beta^2
\end{equation}
where $F_N= M_1M_2/(c_W^2 M_1+ s_W^2 M_2)+ M_Z^2 c_{2\beta}/\mu$.
Including 1-loop corrections, the neutrino mass matrix is written as
 \bea \label{neutrinoloop}
M^{\nu}_{ij}=-\frac{M_Z^2}{F_N} \xi_i\xi_j \cos^2\beta
-\frac{M_Z^2}{F_N} \left( \xi_i \delta_j +\delta_i \xi_j \right)
\cos\beta +\Pi_{ij},
 \eea
where
$\Pi _{ij} $ denotes the 1-loop contribution of the neutrino
self energy, and
\bea \label{comentary}
\delta_i &=& \Pi_{\nu_i \widetilde{B}^0} \left(
\frac{-M_2\sin^2\theta_W}{M_{\widetilde{\gamma}} M_W \tan\theta_W}
\right) + \Pi_{\nu_i \widetilde{W}_3} \left( \frac{M_1
\cos^2\theta_W}{M_{\widetilde{\gamma}}M_W} \right) \nonumber \\
&&+\Pi_{\nu_i \widetilde{H}_1^0} \left( \frac{\sin\beta}{\mu}
\right) +\Pi_{\nu_i \widetilde{H}_2^0}\left(
\frac{-\cos\beta}{\mu} \right),
 \eea

\bea
M_{\tilde{\gamma}} = c_W^2 M_1 + s_W^2 M_2 .
\eea
Exact  expressions of  the 1-loop contributions $\Pi's$ are
presented in Ref.~\cite{phdjung}.
We note that the conventional tau-stau,
bottom-sbottom loops and neutral scalar (sneutrino/neutral Higgs
boson) loops are essential to achieve  a realistic neutrino
mass matrix for $\tan\beta \lesssim 30$, and they are given by
\begin{eqnarray} \label{Mloop}
 M^{loop}_{ij}= &&3{\lambda'_{i}\lambda'_{j}\over8\pi^2}
{m_b^2(A_b+\mu\tan\beta) \over
m_{\tilde{b}_1}^2-m_{\tilde{b}_2}^2} \ln{m_{\tilde{b}_1}^2\over
m_{\tilde{b}_2}^2}   \nonumber \\
&+&~ {\lambda_{i}\lambda_{j}\over8\pi^2}
 {m_\tau^2(A_\tau+\mu\tan\beta) \over
m_{\tilde{\tau}_1}^2-m_{\tilde{\tau}_2}^2}
\ln{m_{\tilde{\tau}_1}^2\over m_{\tilde{\tau}_2}^2}\,.
\end{eqnarray}
It is found that the charged scalar (slepton/charged Higgs boson)
loops  are important only for $tan\beta >30$ as observed in
Ref.~\cite{valle}, whereas  the neutral scalar loops are 
important for all $\tan\beta$ in our case.
We have observed that the tree level masses are  much larger than
 the  1-loop contributions in mSUGRA, which makes it difficult to accommodate
the solar and atmospheric neutrino oscillations.
Thus, we need some cancellation in $\xi_i$  so that the
tree mass becomes comparable to the 1-loop contributions. Then,
we can obtain a desirable neutrino mass matrix the by combining
the tree and 1-loop masses appropriately, but it is possible
in some fine-tuned parameter space, as will be shown later. 
In this case, however, we lose nice predictability of atmospheric and solar
neutrino mixing angles measurable in colliders.

\section{Numerical results: Fitting the neutrino data}

\begin{figure}\label{ratio}
 \epsfig{file=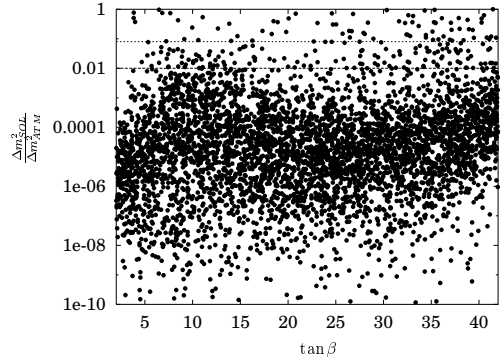,height=5.2cm,width=7.0cm} \caption{Plots
 $\tan \beta$  vs. mass square ratio  for general points, the
 region between the two straight lines is the allowed one by
 neutrino data}
 \end{figure}

\begin{center}
\begin{figure*}
\epsfig{file=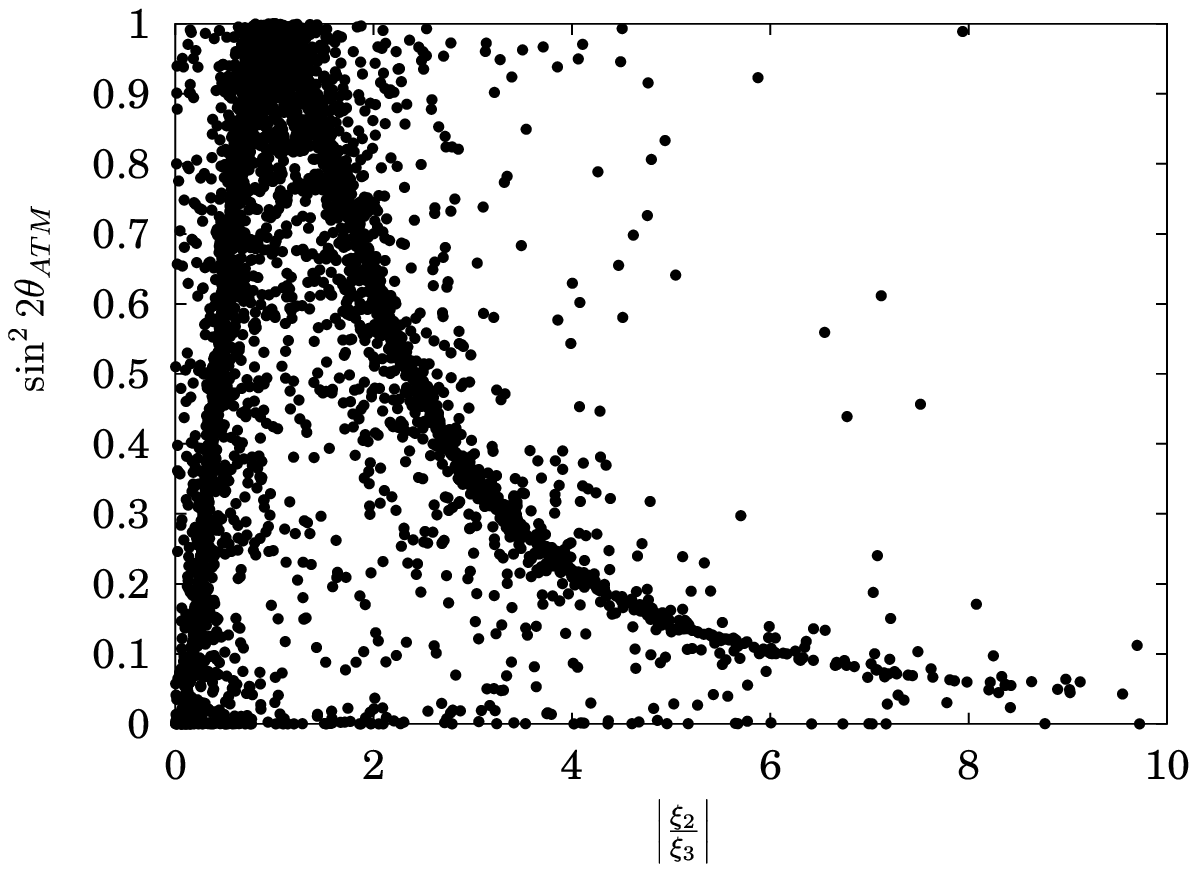,height=4.5cm,width=5.0cm}
\epsfig{file=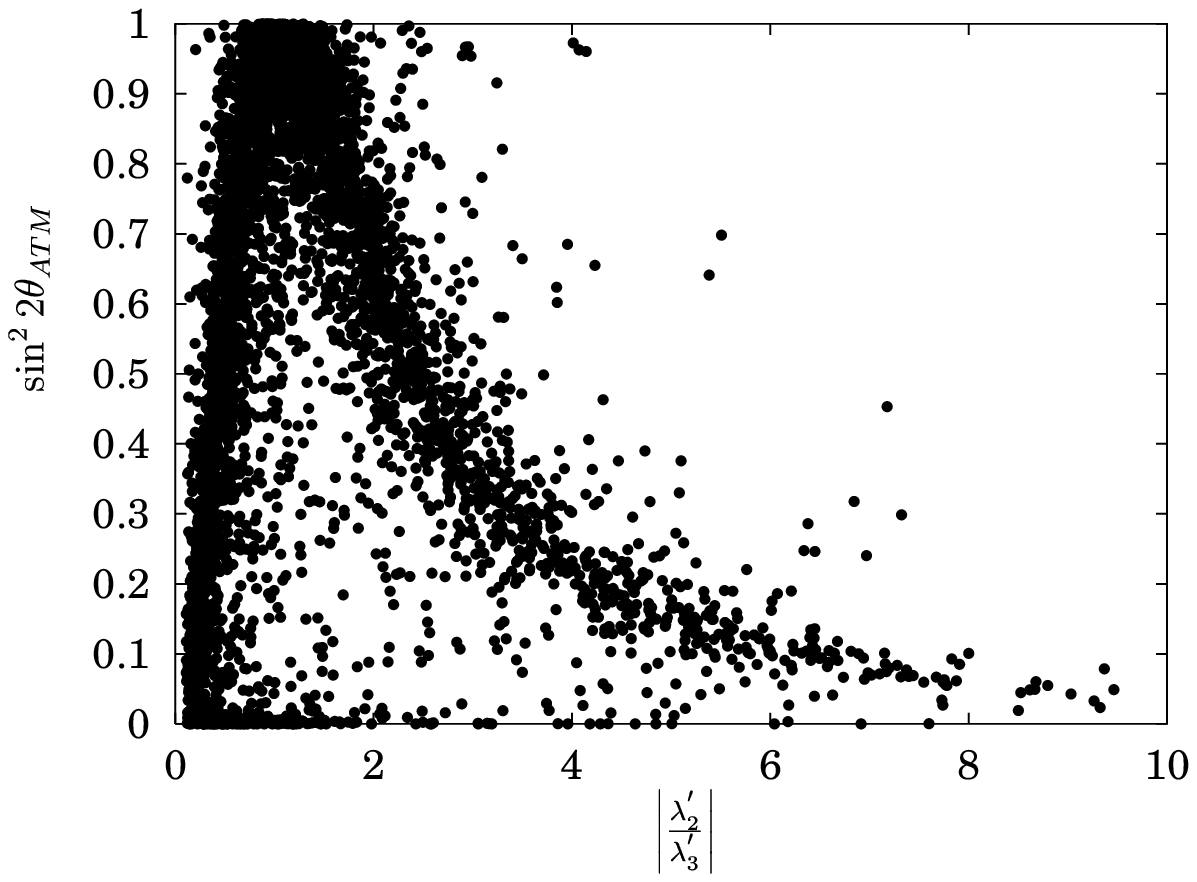,height=4.5cm,width=5.0cm}
\caption{Atmospheric neutrino mixing angle vs. $ \left|
\xi_3/\xi_3 \right| $ and $ \left| \lambda_2^{'} / \lambda_3^{'}
\right| $ for general points }
\end{figure*}
\end{center}

\begin{center}
\begin{figure*}
\epsfig{file=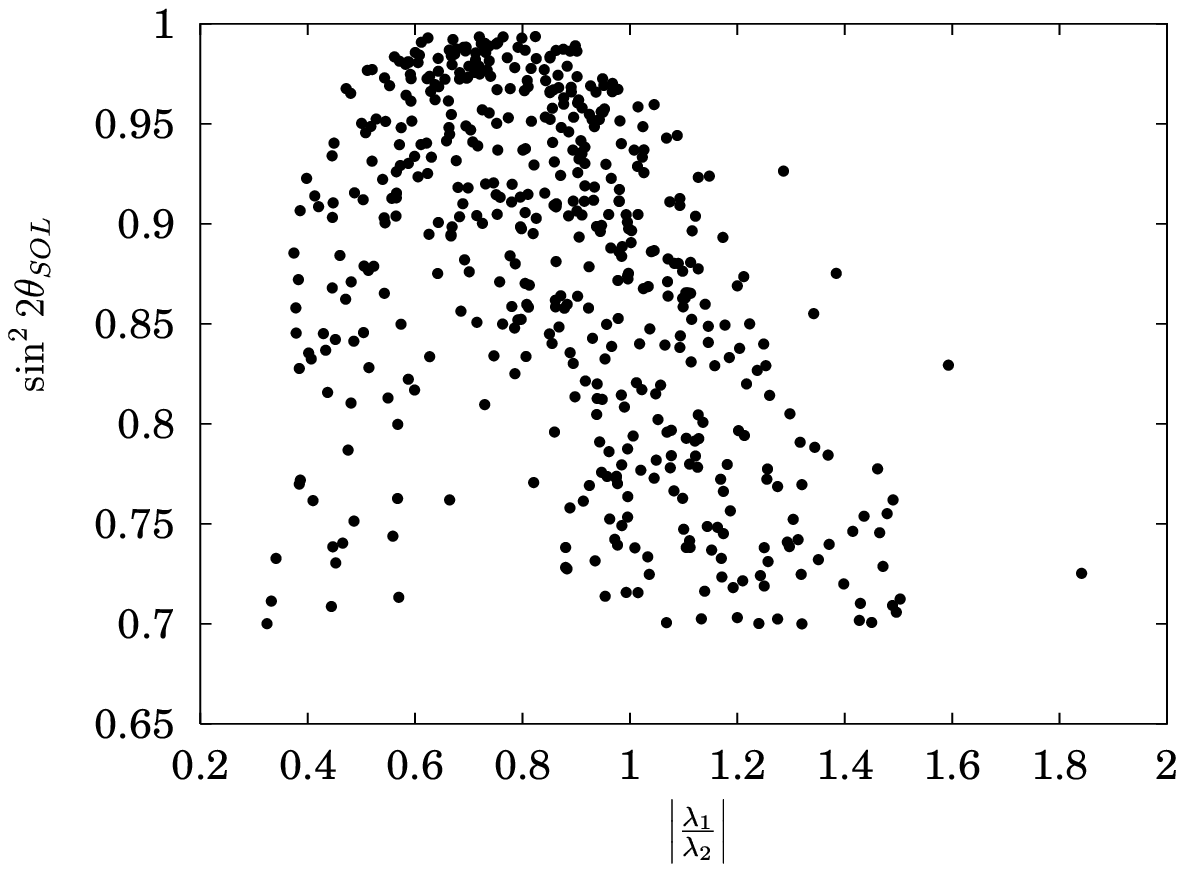,height=4.5cm,width=5.0cm}
\epsfig{file=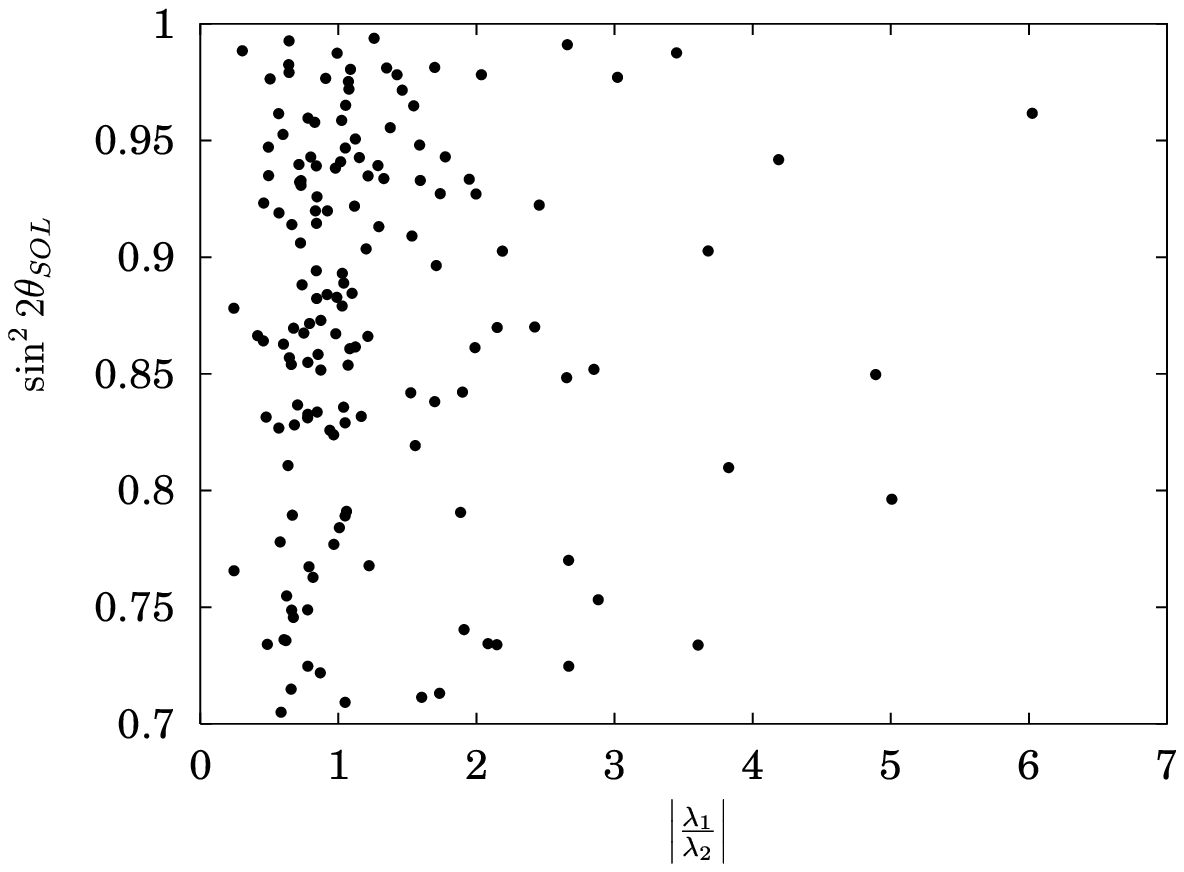,height=4.5cm,width=5.0cm}
\caption{The solar mixing angle and $\lambda_1/\lambda_2$ ratio
for solution points, Left~:~small $\tan \beta, \sim 3 -
15$~Right~:~for large $\tan \beta, \sim 30 - 40$}
\end{figure*}
\end{center}

%\begin{center}

\vspace{-1.5cm}

First of all, we obtain the parameter sets consistent with the solar and
atmospheric neutrino data by scanning the input
parameters  within the ranges,
  \bea
 100 {\rm GeV} \leq &m_0& \leq 1000 {\rm GeV}, \\
 100 {\rm GeV} \leq &M_{1/2}& \leq 1000 {\rm GeV}, \\
 0 {\rm GeV} \leq &A_0& \leq 700 {\rm GeV}, \\
 2 \leq &\tan \beta& \leq 43 .
 \eea
The ranges for $R-$parity  violating parameters we scan are given
 \bea
4 \times 10^{-6} \leq & \left| \lambda_1 \right|& \leq 6 \times
10^{-4}, \\
4 \times 10^{-6} \leq & \left| \lambda_2 \right|&\leq 6 \times
10^{-4}, \\
3 \times 10^{-9} \leq & \left| \lambda_1^{'} \right|& \leq
10^{-4}, \\
4 \times 10^{-6} \leq & \left| \lambda_2^{'} \right|& \leq
10^{-3}, \\
4 \times 10^{-6} \leq & \left| \lambda_3^{'} \right|& \leq
10^{-3}. \eea
We set the signs
of $ M_{1/2}$ and $A_0$ arbitrary, but find that most
of the allowed  parameter space  corresponds to the case that both signs are
positive.
We also find that there are strong correlations between atmospheric neutrino 
mixing angle and $\lambda^{'}_i$'s. 
In order to account for the large atmospheric angle and
small Chooz angle, we should take $ \lambda_1^{'} \ll
\lambda_2^{'} \approx \lambda_3^{'}$. We have  confirmed this by
numerical calculations. In mSUGRA scenario, the tree level value
of neutrino mass is much larger the 1-loop contributions in
general, whereas the allowed parameter sets from
the neutrino data are not tree-dominant at all. So we should
make the tree mass terms comparable to the 1-loop contributions, and
it can be possible in the case that some cancellation in $xi_i$ occurs
%the
%total sum of the tree and loop contributions has to be arranged to
%produce a right neutrino mass matrix, in some fine-tuned parameter
%space. In other words, a cancellation in $\xi_i$ is needed to
%reduce the tree mass so that it becomes comparable to the loop mass. 
In doing so, however, we lose beautiful predictability of atmospheric 
and solar neutrino mixing angles measurable in colliders.
FIG.1 shows  scattered  plot of the ratio $\Delta m^2_{sol}/\Delta
m^2_{atm}$  {vs.} $\tan\beta $ in the randomly generated parameter space. 
The points in the region between the two dotted lines
give the right value of the mass square ratio. Only small set of scattered
points lie in that region.  We find that only small number of points among them
are consistent with bi-large mixing.
%If we require bi-large mixing,
%the number of points is reduced drastically. 
%\begin{center}
%\begin{figure}[]
%\epsfig{file=delm3-gen.eps,height=7.5cm,width=7.5cm}
%\epsfig{file=delm3-sol.eps,height=7.5cm,width=7.5cm}
% \caption{Plot $\lambda_2$
% vs. $|(m_3^{loop}-m_3^{tree})/m_3^{tree}|$,  Left: for general
%points,  Right: for points satisfying the neutrino masses and
%mixing}
%\end{figure}
%\end{center}
%\begin{center}
%\begin{figure}[]
%\epsfig{file=xieta-sol.eps,height=7.5cm,width=7.5cm} \caption{The
%deviation of $\eta_i$ from $\xi_i$ direction, for solution points}
%\end{figure}
%\end{center}
%
%\begin{center}
%\begin{figure}
%\epsfig{file=delxi-small.eps,height=7.5cm,width=7.5cm}
%\epsfig{file=delxi-big.eps,height=7.5cm,width=7.5cm} \caption{The
%effect of svev 1-loop correction, Left~:~ for $\tan \beta = 3 -
%15$, Right~:~for $\tan \beta = 30 - 40 $~ ,both for solution
%points}
%\end{figure}
%\end{center}
 
We note that the relative sizes of R-parity violating
parameters are correlated with neutrino mixing pattern.
But due to the cancellation in $xi_i$, it turns out that such a correlation is 
rather weak.
In FIG. 2, we see some correlations between the atmospheric neutrino mixing 
angle and $\xi_2/\xi_3$, 
$\lambda'_2/\lambda'_3$ for general parameter sets although
they are not very strong because many data points do not lead to tree-dominant
neutrino masses.
Even though the correlations look so weak, we can obtain some
constraints on $\left|\lambda'_2/\lambda'_3 \right|$ from the neutrino
data $
0.4 \lesssim |\lambda'_2/\lambda'_3| \lesssim 2.5 $ for small $\tan\beta$ and
$0.3 \lesssim |\lambda'_2/\lambda'_3| \lesssim 3.3 $ for large $\tan\beta$
In FIG. 3, we plot the solar mixing angle {\it vs.} 
$\lambda_1/\lambda_2$ ratios for small $\tan\beta$ and
large $\tan\beta$ cases, respectively.
We see that  there are  strong correlations between the solar mixing angle and
$\lambda_1/\lambda_2$ ratios. Similar to the above case,
we can get some constraints 
 $ 0.3 \lesssim |\lambda_2/\lambda_3| \lesssim 1.6$ for small $\tan\beta$ and
$ 0.2 \lesssim |\lambda_2/\lambda_3| \lesssim 5$ for large $\tan\beta$.
Collecting these constraints, the solutions are found for the
ranges of
 \bea
 \lambda_{1,2}, \lambda'_{2,3} &=& (0.1-2)\times 10^{-4}, \\
\lambda'_1 &<& 2.5\times 10^{-5}.
 \eea

 Let's summarize the results of this section. In general, in large
 portion of parameter space, the tree level contribution is much
 larger than those from 1-loop, so it is hard to get the right
 mass square ratio which is about ${\cal O}(10^{-2})$. The solutions
 are possible by suppressing the tree level contributions, which
 arises through the cancellation in $\xi_i$. In this case, the
 correlations between $\xi_i$ and mixing angles become worse. In
 addition to the important 1 loop correction by stau  and sbottom,
 the neutral scalar loop can give the important contribution to
 the neutrino masses, through the deviation of $\eta_i$ from the
 direction of $\xi_i$, which determine the tree level neutrino
 mass matrix. Instead of those, there are stronger correlations
 between the atmospheric mixing angle and $\lambda^{'}_2/\lambda^{'}_3$,
 the solar mixing angle and $\lambda_1/\lambda_2$. For small $\tan
 \beta$, it is possible to probe these in collider signal. It will
 be shown in the next section.
%\begin{center}
%\begin{figure}
%\epsfig{file=xi23atm-small.eps,height=7.5cm,width=7.5cm}
%\epsfig{file=lamp23atm-small.eps,height=7.5cm,width=7.5cm}
%\caption{For solution points, small $\tan \beta, \sim 3 - 15$ }
%\end{figure}
%\end{center}
%\begin{center}
%\begin{figure*}
%\epsfig{file=xi23atm-big.eps,height=7.5cm,width=7.5cm}
%\epsfig{file=lamp23atm-big.eps,height=7.5cm,width=7.5cm}
%\caption{For solution points, large $\tan \beta, \sim 30 - 40$}
%\end{figure*}
%\end{center}
%\section{Collider signals of the model}

\begin{table*}[t]
\begin{center}
\begin{tabular}{c|ccc}
\hline
 & & setA : Stau LSP &  \\
\hline \hline
 &  $\tan\beta=5.15$ &  $ {\rm sgn} (\mu) = -1$ &  $\mu= -801.23$ GeV    \\
 &  $ A_0 = 39.47 $GeV  &$ m_0 = 105.02  $ GeV& $M_{1/2} = 671.02$ GeV   \\
\hline
 $\tilde{\lambda}'_i$   &
$8.403\times10^{-6}$ & $-7.663\times10^{-5}$ & $-6.792\times10^{-5}$ \\
$\tilde{\lambda}_i$   &
   $8.739\times10^{-5}$ & $-7.444\times10^{-5}$ & 0     \\
$\xi^0_i$ &
   $-1.003\times10^{-6}$& $-3.642\times10^{-6}$ & $-4.401\times10^{-6}$ \\
$\xi_i$ &
   $-9.897\times10^{-7}$& $-3.045\times10^{-6}$ & $-3.728\times10^{-6}$ \\
$\eta_i$ &
   $1.011\times10^{-6}$& $-1.313\times10^{-5}$ & $-1.192\times10^{-5}$ \\
\hline
BR  &  $e$  & $\mu$   & $\tau$   \\
\hline
$l_i \nu$ & 28.93 \%  &  20.99 \% & 50.02 \% \\
$ \bar{t}~ b$ & & $\sim$  0.066 \% & \\
\hline
       & $m_{\tilde{\tau}_1}$=278.59 GeV
     & \hfill $\Gamma=$& $2.921 \times 10^{-7}$ GeV  \\
\hline \hline
 &  \hspace{1.8cm}$\sigma_{e^+ e^- \rightarrow \tilde{\tau}_1 \tilde{\tau}_1^*}$
&$\simeq$ 1.450 $\times 10^{-2}$ (Pb), & $\sqrt{s}$ = 1 TeV  \\
\hline
\end{tabular}
\end{center}
$$ \begin{array}{l}
 (\Delta m^2_{31},~ \Delta m^2_{21})=
     (2.50\times10^{-3},~1.13\times10^{-4})~ \mbox{eV}^2  \cr
 (\sin^22\theta_{atm},~ \sin^22\theta_{sol},~ \sin^22\theta_{chooz})
 =(0.98,~0.77,~0.03)
\end{array} $$
\caption{A trilinear model realizing the LMA solution for stau-LSP
case. Here the couplings $\tilde{\lambda}'_i$ and
$\tilde{\lambda}_i$ can be considered as input parameters defined
at the weak scale. }
\end{table*}

\begin{table*}[]
\begin{center}
\begin{tabular}{c|ccc}
\hline
 & & setB : Neutralino LSP &  \\
\hline \hline
 &  $\tan\beta=4.94$ &  $ {\rm sgn} (\mu) = -1$ & $\mu= -200.46$ GeV    \\
 &  $ A_0 = 38.93 $GeV  &$ m_0 = 333.66  $ GeV& $M_{1/2} = 160$ GeV   \\
\hline
 $\tilde{\lambda}'_i$   &
$-9.326\times10^{-9}$ & $-7.811\times10^{-5}$ & $-7.560\times10^{-5}$ \\
$\tilde{\lambda}_i$   &
   $-5.628\times10^{-5}$ & $-7.345\times10^{-5}$ & 0     \\
$\xi^0_i$ &
   $-1.234\times10^{-6}$& $3.247\times10^{-7}$ & $1.880\times10^{-6}$ \\
$\xi_i$ &
   $-1.211\times10^{-6}$& $2.749\times10^{-7}$ & $1.831\times10^{-6}$ \\
$\eta_i$ &
   $-2.007\times10^{-6}$& $-1.169\times10^{-5}$ & $-8.742\times10^{-6}$ \\
\hline
BR  &  $e$  & $\mu$   & $\tau$   \\
\hline
$\nu jj$ &  &47.01 \%   &  \\
$ l_i^\pm jj$ &3.88 $\times 10^{-2}$ \% & 2.00 $\times 10^{-3}$\% & 8.87 $\times 10^{-2}$\% \\
$ \nu l_i^\pm l_3^\mp$  & 9.76 \% & 16.60 \%&  26.39 \% \\
\hline
       & $m_{\tilde{\chi}^0_1}$=59.37 GeV
     & \hfill $\Gamma=$& $7.137 \times 10^{-15}$ GeV  \\
\hline \hline
 &  \hspace{1.8cm}$\sigma_{e^+ e^- \rightarrow \tilde{\chi}^0_1 \tilde{\chi}^0_1}$
&$\simeq$ 4.897 $\times 10^{-2}$ (Pb), & $\sqrt{s}$ = 1 TeV  \\
\hline
\end{tabular}
\end{center}
$$ \begin{array}{l}
 (\Delta m^2_{31},~ \Delta m^2_{21})=
     (2.51\times10^{-3},~9.64\times10^{-5})~ \mbox{eV}^2  \cr
 (\sin^22\theta_{atm},~ \sin^22\theta_{sol},~ \sin^22\theta_{chooz})
 =(0.98,~0.99,~0.008)
\end{array} $$
\caption{Neutralino-LSP case, only 3-body decays are possible}
\end{table*}

\begin{table*}[]
\begin{center}
\begin{tabular}{c|ccc}
\hline
 & & setC : Neutralino LSP &  \\
\hline \hline
 &  $\tan\beta=4.05$ &  $ {\rm sgn} (\mu) = -1$ &$\mu=-535.92$ GeV     \\
 &  $ A_0 = 14.28 $GeV  &$ m_0 = 113.88  $ GeV& $M_{1/2} = 444.57$ GeV   \\
\hline $\tilde{\lambda}'_i$   &
$-9.069\times10^{-9}$ & $-1.396\times10^{-4}$ & $-1.842\times10^{-4}$ \\
$\tilde{\lambda}_i$   &
$-7.966\times10^{-5}$ & $-7.995\times10^{-5}$ & 0     \\
$\xi^0_i$ &
$1.274\times0^{-6}$& $2.896\times10^{-6}$ & $2.138\times10^{-6}$ \\
$\xi_i$ &
$1.212\times10^{-6}$& $2.976\times10^{-6}$ & $2.409\times10^{-6}$ \\
$\eta_i$ &
$4.038\times10^{-7}$& $-1.305\times10^{-5}$ & $-1.768\times10^{-5}$ \\
\hline
BR  &  $e$  & $\mu$   & $\tau$   \\
\hline
$\nu jj$ &  &14.77\%   &  \\
$ l_i^\pm jj$ &6.36$\times 10^{-2}$ \% & 3.84 $\times 10^{-1}$\% & 2.51 $\times 10^{-1}$\% \\
$ \nu l_i^\pm l_3^\mp$  & 20.96 \% & 21.186
\%& 42.09 \% \\
\hline & $m_{\tilde{\chi}^0_1}$=195.00 GeV
& \hfill $\Gamma=$& $4.744\times 10^{-11}$ GeV  \\
\hline \hline &  \hspace{1.8cm}$\sigma_{e^+ e^- \rightarrow
\tilde{\chi}^0_1 \tilde{\chi}^0_1}$
&$\simeq$ 0.197 (Pb), & $\sqrt{s}$ = 1 TeV  \\
\hline
\end{tabular}
\end{center}
$$ \begin{array}{l}
(\Delta m^2_{31},~ \Delta m^2_{21})=
(2.50\times10^{-3},~6.81\times10^{-5})~ \mbox{eV}^2  \cr
(\sin^22\theta_{atm},~ \sin^22\theta_{sol},~
\sin^22\theta_{chooz}) =(0.95,~0.96,~0.007)
\end{array} $$
\caption{Neutralino-LSP case, both 2- and 3-body decays are
possible.}
\end{table*}

\section{Collider signals of the model}
 It is expected that the luminosity of the next linear collider
can reach the $1000 {\rm fb}^{-1} /{\rm yr}$, and the center of
mass energy over $1~{\rm TeV}$\cite{desy1}\cite{desy2}. With such a
capacity of the linear collider, it could be possible to  prove
the supersymmetric particles
pair produced\cite{bartl} as well as the decay of LSP
inside the detector\cite{Chun:2002vp}. We present here the total
cross section and decay rate for various parameter sets, and
discuss the possibility to probe the structure of R-parity
violating parameters constrained by neutrio data. In the previous
sections we got the parameter sets which are consistent with the known
neutrino data. Those results suffice to  determine which is the LSP,
neutralino or light stau. In TABLE I-III, we present the results of
decay rate and branching ratios. The
TABLE I shows the results for the stau LSP case with small $\tan
\beta$.  TABLE II and III correspond to the Neutralino LSP cases for
small $\tan \beta$.  
The former is the case that only 3-body decay is permitted,
whereas the latter is the case that both 2-body and 3-body decays are permitted.
Since the decay lengths are smaller than a few $cm$ for all cases, 
it is possible to detect their decay modes in the next linear collider.

Firs of all, let's discuss the stau LSP cases. For small $\tan \beta$,
$\tilde{\tau}_1 \sim \tilde{\tau}_R$ due to the small off-diagonal part. 
Then the light stau  almost decays into leptons via
$\lambda_i L_i L_3 E^c_3$ terms in the superpotential. In this
case, the following relation holds,
 \bea
 Br(e\nu) : Br(\mu \nu) : Br(\tau \nu)&& \nonumber \\
 && \hspace{-2cm} \simeq
|\lambda_1 |^2 : |\lambda_2 |^2:  |\lambda_1 |^2 + |\lambda_2 |^2.
 \eea
Thus, by observing the lepton  branching ratios, one can measure the
ratio of $\lambda_1 $ and $\lambda_2$. If this value is out of the
range constrained by neutrino data, given in the last section, we
can exclude this model. If $ \tan \beta$ become larger, the above
relation does not hold any more because the Yukawa coupling plays
an important role in this case. In other words, $\tilde{\tau}_1 $
cannot be considered as almost $\tilde{\tau}_R$, but should
be the combination of $\tilde{\tau}_L$ and $\tilde{\tau}_R$ and then
the couplings are given by the mixture of $ \lambda_i$ and $h_\tau$.

 Next, let's consider the neutralino LSP cases. The 2-body decay
rates are proportional to $\left| \xi_i \right|^2$,
 \bea
&&\Gamma (\nu_i Z) \propto |\xi_i|^2 ,\\
&&\Gamma (l_i W) \propto |\xi_i|^2.
 \eea
For 3-body decay modes, the lepton-quark-qurak branching ratios
are proportional to $\left| \xi_i \right|^2$ since $W$ boson
exchange diagrams are dominant. So we can have the following
relation, \bea Br(ejj) : Br(\mu jj) : Br( \tau jj) = |\xi_1|^2 :
|\xi_2|^2 : |\xi_3|^2. \eea  
Thus, we can obtain the information of $\xi_i$ ratio by measuring
the branching ratios, but
it is difficult to test mSUGRA scenario for massive neutrinos
via collider signals because the correlation between
$\xi_i$'s and  neutrino oscillation parameters is diminished
as pointed out in the last section.
Similar to the stau LSP case, for small $\tan \beta$, $\nu l_i^\pm
\tau^\mp$ branching ratios provide $\lambda_i$ information on them;
\bea && Br(\nu e^{\pm} \tau^{\mp}) : Br( \nu \mu^{\pm} \tau^{\mp})
: Br(\nu \tau^{\pm} \tau^{\mp}) \nonumber \\
 &&\simeq |\lambda_1 |^2 :
|\lambda_2 |^2:  |\lambda_1 |^2 + |\lambda_2 |^2
\label{ccc} \eea
 Likewise, if we measure the above branching ratios,
the models can be tested by comparing them with the range allowed
by neutrino data, 
But since, for large $\tan \beta$, the terms generated from
$h_\tau L_3 H_1 E_3^c$ ~in the superpotential become large, and thus 
the above relation (\ref{ccc}) breaks, it is impossible to test this scenario.
 The examples are presented in the TABLE II and
III. In addition, we  note that $\eta_i$'s are rather big in all cases, which
implies that some cancellations in $\xi_i$'s occur, as
expected before.

\section{Conclusion}

Based on the recent precise experimental results for neutrino
oscillations, 
%As a result of many experiments, we now have very precise data
%sets on neutrino masses and mixing angles; large but not maximal
%solar neutrino mixing angle, almost maximal atmospheric neutrino
%angle, very small reactor angle, and corresponding mass square
%ratio\cite{Fukuda:1998mi}. To explain this neutrino
%oscillation, it is good to use the minimal supergravity model with 
%R-parity violation, which has lepton number violation
%interactions generically\cite{Hall:1983id}. 
we have studied the minimal supergravity model with the lepton number violation
via R-parity violating terms \cite{Hall:1983id}. 
Since it is impossible to  account for large mixing angle solution
of solar neutrino problem in mSUGRA with  only bilinear R-parity violations, 
we should include the general trilinear R-parity violations. 
So there are additional 5 parameters $\lambda_i$
and $\lambda_i^{'}$ besides the 5 parameters which are from
the ordinary minimal supergravity model. 
%If the future linear collider is
%settle down, it is expected to grow
In future collider which can reach its CM energy over 1 TeV and
luminosity up to 1000 ${\rm fb}^{-1}$ in a year\cite{desy1}\cite{desy2}, 
we expect that it is possible to probe mSUGRA without R-parity  via
the LSP decay.
We have presented how neutrino mass
matrix from mSUGRA with R-parity violation can be constructed at the 1-loop 
level. From our analysis, we found that most parameter space which is consistent with neutrino data leads to  1-loop contributions as large as comparable
with tree level masses.
We have found the
strong correlation between solar neutrino mixing angle and
$\lambda_1 / \lambda_2 $, and also between atmospheric neutrino
mixing angle and $ \lambda_2^{'} / \lambda_3^{'} $. 
For searching the LSP decays, we have considered two cases in mSUGRA,
neutralino LSP and scalar tau LSP, and
searched each parameter space regions by varying $m_0$ and
$M_{1/2}$. For our purpose, we have calculated the production
cross section for each cases. For stau LSP case, 2-body decay
modes are dominant. Since stau LSP is almost
$\tilde{\tau}_R$ for most parameter space,  its branching ratios of the decay 
into top and bottom quark is negligibly small.
Since stau LSP decay into $ l_i
\nu$ takes place via $\lambda_i$ couplings, it is possible to obtain the
information on $ \lambda_i$'s from branching fractions. We found
that $ Br(e \nu) : Br(\mu \nu) : Br(\tau \nu) = \left|\lambda_1
\right|^2 : \left| \lambda_2 \right|^2 : \left|\lambda_1 \right|^2
+ \left| \lambda_2 \right|^2 $ for not too large $\tan \beta$. For
neutralino LSP cases, similar analysis has been done. Unlike the stau LSP
cases, neutralino LSP can decay into 3-body final states as well as
2-body decay. 
>From the branching fractions
$Br( l_i jj)$ for various decay channels of neutralino LSP, 
we have determined the ratios of $\xi_i$'s, which in turn
give the desirable tree level neutrino mass matrix. 
But, due to rather large loop contributions to neutrino masses
the predictability for neutrino parameters is diminished.
For small $\tan \beta $, we have obtained the
$\lambda_i$'s information which are strongly correlated with solar
neutrino mixing angle from the branching fractions $Br(\nu l_i^\pm
\tau^\mp)$. But unluckily, since Yukawa
couplings become larger for large $\tan\beta$, 
it is impossible to get information on $
\lambda_i$'s. Consequently, for both stau LSP and neutralino LSP,
it is possible to probe the relations between neutrino oscillation
and R-parity violating couplings from the branching fractions. But
it becomes hard for large $\tan \beta$. Thus,
any observation of the sever deviation from
these branching fractions in the future collider may exclude
scenarios for neutrino masses in the framework of mSUGRA with
appropriate R-parity violation.

\acknowledgements{SKK is supported by BK21 program of the Minstry 
of Education in Korea, and DWJ is supported by KRF-2002-070-C00022.}

%\end{multicols}

\end{document}